\newcommand{\beq}{\begin{equation}}
\newcommand{\eeq}{\end{equation}}
\newcommand{\beqns}{\begin{equation}}
\newcommand{\eeqns}{\end{equation}}
\newcommand{\beqar}{\begin{eqnarray}}
\newcommand{\bs}{\begin{eqnarray*}}
\newcommand{\eeqar}{\end{eqnarray}}
\newcommand{\es}{\end{eqnarray*}}
\newcommand{\beqml}{\begin{mathletters}}
\newcommand{\eeqml}{\end{mathletters}}
\newcommand{\Ghat}{\hat{G}}

\documentstyle[aps,preprint]{revtex}
\begin{document}
\draft
\title{Iterated Function System and Diffusion in the
Presence of Disorder and Traps}
\author{Thomas Wichmann, Achille Giacometti
and K.~P.~N. Murthy \cite{on leave}}

\address{Institut f\"{u}r Festk\"{o}rperforschung
des Forschungszentrums J\"{u}lich \\
Postfach 1913, D-52425,  J\"{u}lich, Germany}
\date{\today}
\maketitle

\begin{abstract}
The escape probability $\xi_{x}$
from a site $x$ of a  one-dimensional disordered
lattice with trapping is treated as a discrete dynamical evolution
by random iterations over nonlinear maps parametrized by the
right and left jump probabilities. The invariant measure
of the dynamics is found to be a multifractal. However the
measure becomes uniform over the support when the
disorder becomes weak for any non-zero trapping probability.

Implications of our findings
in terms of diffusion are discussed.
\end{abstract}
%
\newpage
\narrowtext
Diffusion in the presence of disorder and trapping has become by
now a classic field.  This is mainly because the problem
{\it per se} is mathematically interesting and is well posed;
also it has a tremendous potential for a wide range of applications
which include migrations of optical excitations \cite{AK},
polymer physics \cite{de Gennes} and diffusion-limited binary
reactions \cite{Mik}. See e.g. \cite{Reviews} for an exhaustive review.

The standard approach to this class of problems is to write
down a second order master equation
for the probability of the particle to be
at a lattice site at a given time,
and solve it employing analytical or numerical techniques,
see for example
\cite{Alexander et al}. An alternate approach,
based on the first passage time (FPT) formulation, has attracted
growing attention in the recent times \cite{MK,KM,MRK,VDB,MVI}.
This approach has an advantage in that
the master equation is first-order to start with. All
the transport properties of the system can also be calculated
from the first passage time formulation.

Employing FPT formulation for the Sinai model \cite{Sinai}
it was recently shown that the distribution of the
mean FPT over the disorder exhibits interesting multifractal
scaling \cite{MRK}.
Also the probability to escape from
one site of the lattice to the next was
found to have self similar fluctuations \cite{VDB,MVI}.
The Sinai model however, is an  highly idealized,
albeit interesting, mathematical model, and whose link
to physical reality appears to
be rather abstract.

In this work we shall consider a more realistic model for
diffusion where a particle diffuses by overcoming random barriers
but can also be trapped at various sites with site  dependent
random probabilities \cite{G}.
The main characteristic of this
class of models is that the total probability (called the survival
probability) is {\it not} a conserved quantity. It has been shown
that the survival probability is an highly fluctuating function
of time and these fluctuations lead to interesting and unexpected
behaviour  like enhanced diffusion, breaking of
self-averaging and emergence of Lifshitz tails \cite{G et al}.

We shall show that this model, when disorder is strong leads to
self-similar fluctuations
of the escape probability, and these can be characterised
employing multifractal formalisms.
However this feature of multifractality disappears
when there is no trapping, what ever may be the strength of disorder.
More importantly, when the strength of disorder goes to zero
the multifractality disappears even with arbitrary non-zero trapping
probability. Purely from  methodological
point of view, we connect diffusion
in a trapping environment to an Iterated Function System (IFS)
\cite{Barnsley}. Such a formulation, connecting random walks
and iterated function systems,
was proposed very recently in the context of a binary model for Sinai disorder
\cite{VDB,MVI}, where we have two maps for random iterations.
Here we extend the formulation to problems
of diffusion on a disordered lattice in the presence
of trapping, where we have infinity of maps parametrized by
the jump probabilities which are chosen randomly from
a well specified distribution that models the disorder.
We restrict our attention to a one-dimensional lattice since,
as often the case \cite{LM}, it is contains all the essential
characteristics of the higher dimensional systems.
Furthermore one-dimensional systems are amenable to
relatively easy analytical and numerical work.

Let us consider the master equation for the probability
$\Ghat_{x,x+1}(n)$ that a particle makes a first passage
from a site $x$ to a site $x+1$ in $n$ steps on a one-dimensional
lattice of length $N$. At each site $x \ge 1$
we shall indicate by $q_x \in [0,1/2]$ the
probability for making a left jump and by $p_x \in [0,1/2]$
the probability of making a right jump (see Fig \ref{fig1}).
The sojourn probability at site $x$
is given by $\gamma(1-q_x-p_x)$ and the trapping
probability is $(1-\gamma)(1-q_x-p_x)$.
Here $\gamma$ is a parameter which can
be continuously tuned from $0$ (trapping) to $1$ (no trapping)

The master equation for $\Ghat_{x,x+1}(n)$ ($x \ge 1$) then reads:
\beqar \label{FPME}
\Ghat_{x,x+1}(n) &=& p_x \delta_{1,n} +q_x \Ghat_{x-1,x+1}(n-1)+
\gamma(1-q_x-p_x) \Ghat_{x,x+1}(n-1)
\eeqar
with the boundary condition that site $x=-1$ is perfectly reflecting:
\beqar \label{FPBC}
\Ghat_{0,1}(n) &=& p_0 \delta_{1,n} + \gamma(1-p_0) \Ghat_{0,1}(n-1)
\eeqar
and that $\Ghat_{x,x+1}(n)=0$ for $x \le -1$.
We assume that $\{q_x, \; x=1,\; N-1; \;  p_x, \; x=0, \; N-1 \}$
constitute a set
of independent random variables identically distributed in the
range $(0-1/2)$, and the common distribution is given by
\beqar \label{disorder}
\pi(w) &=& 2^{1-\beta}(1-\beta) w^{-\beta} \theta(w) \theta(1/2-w)
\eeqar
where $w=q,\ p$ and $\beta \in [0, /; 1)$. Here $\theta(\cdot)$ is
the usual Heaviside function. This distribution is known to
produce anomalous diffusion when the disorder is strong
($\beta \to  1$) and there is no trapping \cite{Alexander et al}.
For $\beta = 0$, we find that the distribution is uniform in
the range zero to half. Thus $\beta$ can be tuned from
$0$ (weak disorder) to $1$ (strong disorder).

Equations (\ref{FPME}) and (\ref{FPBC}) are  readly
solved employing generating function technique. We define
\beqar \label{GF}
G_{x,x+1}(z) &=& \sum_{n=0}^{+\infty} z^n \Ghat_{x,x+1}(n)
\eeqar
Upon the use of convolution theorem,
\beqar \label{CT}
G_{x-1,x+1}(z) &=& G_{x-1,x}(z) \; G_{x,x+1}(z)
\eeqar
we get:
\beqar \label{solution bulk}
G_{x,x+1}(z) &=& \frac{z p_x}{1-\gamma z (1-q_x-p_x) -z q_x G_{x-1,x}(z)}
\eeqar
for $x \ge 1$ and
\beqar \label{solution surface}
G_{0,1}(z) &=& \frac{z p_0}{1-\gamma z(1-p_0)}
\eeqar
This solution has earlier been obtained in Ref.\cite{MK}.

We are interested in the behaviour of the {\it escape probability},
namely the total probability for the first passage from $x$ to $x+1$.
This is given by:
\beqar \label{escape probability}
\xi_x &\equiv & G_{x,x+1}(z=1) = \sum_{n=0}^{+\infty} \Ghat_{x,x+1}(n)
\eeqar
Using eqns. (\ref{solution bulk}) and (\ref{solution surface}) it
is immediately seen that the escape probability satisfies
the following one-dimensional recursion:
\beqar \label{map bulk}
\xi_x &=& \frac{p_x}{1-\gamma(1-q_x-p_x)-q_x \xi_{x-1}} \;\; x \ge 1
\eeqar
with the initial condition
\beqar \label{map surface}
\xi_0 &=& \frac{p_0}{1-\gamma(1-p_0)}
\eeqar
Eq. (\ref{map bulk}) can be interpreted as a dynamical map
for a fixed $p$ and $q$.
In fact, since $p$ and $q$ are random, the evolution
$\xi_0 \to \xi_1 \cdots \to \xi_x \to \xi_{x+1}
\to \cdots$ proceeds by random iteration over
the maps parametrized by
$p$ and $q$, which are chosen independently and randomly from
the disorder distribution give by eq. (\ref{disorder}) at each stage
of iteration.
This constitutes an iterated function
system(IFS), see Barnsley \cite{Barnsley}.

It is immediately seen that when $\gamma = 1$, which corresponds to
a lattice with no trapping
eqns. (\ref{map bulk}) and (\ref{map surface})
lead to  $\xi_x=1$, for all $x$
{\it regardless} of the choice of $\{p_x,q_x\}$.

Let us now consider the non-conserved case, for which
$\gamma$ is less than $1$. For given
values of $q$ and $p$ the fixed point of the map (\ref{map bulk}) is
\beqar \label{FP}
\xi^* &=& \frac{[1-\gamma(1-q-p)] - \sqrt{[1-\gamma(1-q-p)]^2 -4pq]}}{2q}
\eeqar
and it is stable. It lies (see Fig \ref{fig2}) in the region delimited by
$\xi^*=0$ corresponding to $p=0$, $q>0$ (only left jump) and
$\xi^*=p/[1-\gamma(1-p)]$ corresponding to $q=0$, $p>0$ (only
right jumps).

We now show that the escape probability exhibits self similar
fluctuations and these can be characterized employing
multifractal formalisms \cite{Feder}.
 From numerical point of view, it proves convenient to
rescale  $\xi_{x}$ in
such a way that the domain is the interval $[0,1]$.
To this end we employ the standard rescaling:
\beqar \label{rescaling}
\frac{\xi -\xi_{\mbox{min}}}
{\xi_{\mbox{Max}}-\xi_{\mbox{min}}}
&\rightarrow& \xi
\eeqar
We denote by $\rho_i(\epsilon)$, the fraction of the total
number of  $\xi$ values that belong to the $i^{th}$ interval
of size $\epsilon=1/N$. Then the
{\it partition function} is given by,
\beqar \label{PF}
Z(Q,\epsilon) &=& \sum_{i=1}^N \rho_i^Q(\epsilon)
\eeqar
where the sum is taken over non-empty intervals only.
We make the following scaling {\it ansatz},
\beqar \label{SA}
Z(Q,\epsilon)\stackrel{\epsilon \to 0}{\sim} \epsilon^{\tau (Q)}
\eeqar
and obtain the scaling exponents $\tau(Q)$ as:
\beqar \label{tau}
\tau(Q) &=& \lim_{\epsilon \rightarrow 0}
\frac{\ln [Z(Q,\epsilon)]}{\ln \epsilon}
\eeqar
Fig. \ref{fig3} shows a log-log plot of $Z(Q,\epsilon)$
versus $\epsilon(=1/N)$ for
$N$ ranging from $10$ to $3 \times 10^6$. The linearity of the curves
establishes unambiguously the scaling {\em ansatz} (\ref{PF}).
 From the scaling exponents we calculate the generalized
Renyi dimensions, given by $D(Q)=\tau (Q)/(Q-1)$.
Legendre transform of $\tau (Q)$, defined as
\beqml
\beqar
f(\alpha) &=& \alpha Q -\tau(Q) \label{legendre:1} \\
\alpha &=&\frac{d}{dQ}\tau(Q) \label{legendre:2}
\eeqar
\eeqml
yields the spectrum of singularities denoted by $f(\alpha )$.

Fig. \ref{fig4}  depicts the scaling exponents $\tau(Q)$. It is well
defined and exhibits clear change in slope, establishing that
the underlying measure is multifractal. Fig. \ref{fig5} depicts the
spectrum of Renyi dimensions $D(Q)$ for
various strengths of disorder($\beta$) and trapping($\gamma$).
We observe that when the disorder is strong ($\beta\to 1$),
$D(Q)$ remains the same for all values of $\gamma \ne 1$.
In other words, the strength of trapping does not influence
the fractal measures of the escape probability, when the disorder
in the lattice is strong. However when the disorder is weak
($\beta\to 0$), the spectrum of Renyi dimensions changes from
one trapping rate to the other. Also in the limit of $\beta\to 0 $,
the $D(Q)$ curve becomes flat with unit intercept, for all
values of $\gamma \ne 1$,
implying that the measure is uniform and
space filling. To capture in a simple fashion the dependence of
the fractal measure on the strength of disorder, we depict in
Fig. \ref{fig6}  the variation of the information dimension $D(1)$ and
the correlation dimension $D(2)$, as a function of $\beta$
for a fixed value of $\gamma = 0$. We find that both
$D(1)$ and $D(2)$ decrease with increasing strengths of disorder.
We plot in Fig. \ref{fig7} the $f-\alpha$ curve.
It is worthwhile noticing that since the slope of $\tau(Q)$ for
$Q \to - \infty$ saturates at unity,
the side of $f(\alpha)$ for $\alpha >1$ does not exhist.

      A natural question that arises in this context
      relates to the implications of our findings
      to the transport properties of the disordered
      systems. More specifically we ask the question:
Is there a connection between anomalous diffusion and fractal
fluctuations?
For example it has already been shown analytically \cite{Alexander et al}
      that even in the absence of trapping, the
      system exhibits anomalous diffusion when the disorder
      is strong ($\beta\to 1$). However when the disorder
      is weak, the anomaly in the diffusion process disappears.
On the other hand, our finding is that when the disorder is strong,
for arbitrary trapping rate, (so long it is non zero),
      the escape probability exhibits multifractal fluctuations.
     However, when the disorder becomes weak ($\beta\to 0$),
      the multifractal features disappear.
     More studies with different disorder models are required
     to shed more light on the issues
     raised here. We hope this work would spur some
     activities along these directions. In any case we
     believe, it would worth the effort to investigate
     the intriguing connection between anomalous diffusion
     and fractal fluctuations in random and trapping environments.

It is a pleasure to thank K. W. Kehr for illuminating
discussions and for a critical reading of the manuscript.
KPNM thanks IFF der KFA J\"{u}lich for the award of a visiting
scientist position during January-April 1995. He also thanks
M. C. Valsakumar and R. Indira  for many discussions  and collabotation on
related work.  AG acknowledges support
from the HCM program under contract ERB4001GT932058.

\begin{figure}
\caption{Definition of the hopping probabilities. At site $x$ the
left jump probability is $q_x$ and the right jump probability is
$p_x$. The sojourn probability is $S_x= \gamma(1-q_x-p_x)$ while the
trapping probability is $T_x=(1-\gamma)(1-q_x-p_x)$ where $\gamma \in [0,1]$.}
\label{fig1}
\end{figure}
\begin{figure}
\caption{The map $\xi_x= f(\xi_{x-1})$ for
$p=0.2$, $q=0.1$, $p=q=0.15$ and $p=0.1$, $q=0.2$. Fixed points are
the interections of the maps with the $\xi_x=\xi_{x-1}$ line.}
\label{fig2}
\end{figure}
\begin{figure}
\caption{Behaviour of the partition function ${\it Z}(Q,\epsilon)$
vs. $\epsilon$ in a log-log plot for $\beta=0.3$ and $\gamma=0.99$.
$Q$ varies are from $-5$ (top curve) to $+10$ (bottom curve)
in units of $1$ with $Q \ne 1$.}
\label{fig3}
\end{figure}
\begin{figure}
\caption{Plot of $\tau(Q)$ vs. $Q$.}
\label{fig4}
\end{figure}
\begin{figure}
\caption{Spectrum of the Renyi dimensions $D(Q)$ vs. $Q$.}
\label{fig5}
\end{figure}
\begin{figure}
\caption{The information dimension $D(1)$  and of the correlation
dimension $D(2)$ as a function of the strength of
the disorder $\beta$, for $\gamma=0$.}
\label{fig6}
\end{figure}
\begin{figure}
\caption{The spectrum of singularities for $\beta=0.7$, $\gamma$ arbitrary
and $\beta=0.3$, $\gamma=0$.}
\label{fig7}
\end{figure}
\end{document}